\documentstyle[epsf]{article}

\def\frac#1#2{ { #1 \over #2} }

\def\braketa#1{\langle #1 \rangle}

\def\braketc#1#2#3{\langle #1 | #2 | #3 \rangle}
\def\gtsim{\mathrel{\hbox{\raise0.2ex
    \hbox{$>$}\kern-0.75em\raise-0.9ex\hbox{$\sim$}}}}
\def\ltsim{\mathrel{\hbox{\raise0.2ex
    \hbox{$<$}\kern-0.75em\raise-0.9ex\hbox{$\sim$}}}}
\def\cJ{{\cal J}}

\def\beq{\begin{equation}}
\def\eeq{\end{equation}}
\def\omegaw{\omega_{\rm w}}
\def\rot{{\rm rot}}
\def\defo{{\rm def}}
\def\out{{\rm out}}
\def\in{{\rm in}}

\makeatletter
\def \refcite#1{\newcount\tmpc\tmpc=0\@for\tmpref:=#1\do{%
\ifnum\tmpc=0\ref{\tmpref}\else\ref{\tmpref}\fi\advance\tmpc by 1}%
}
\makeatother

\def \rcite#1{\refcite{#1})}
\def \scite#1{$^{\rcite{#1}}$}

\begin{document}

\parindent=14 truept
\baselineskip 14 truept

\begin{center}
\begin{minipage}{5in}

\centerline{\large\bf Microscopic Study of Wobbling Motions
         in Hf and Lu Nuclei}

\vspace{3mm}

\centerline{
     Yoshifumi R. Shimizu,
     Masayuki Matsuzaki$^{*)}$,
     and Kenichi Matsuyanagi$^{**)}$}
\centerline{ \it
    Department of Physics, Graduate School of Sciences,}
\centerline{ \it
	Kyushu University, Fukuoka 812-8581, Japan }
\centerline{ \it $^{*)}$
    Department of Physics, Fukuoka University of Education,}
\centerline{ \it
	Munakata, Fukuoka 811-4192, Japan}
\centerline{ \it $^{**)}$
    Department of Physics, Graduate School of Sciences,}
\centerline{ \it
	Kyoto University, Kyoto 606-8502, Japan}

\end{minipage}
\end{center}
\vspace{3mm}

\begin{abstract}
\baselineskip 10 truept

  One of the most striking findings in the recent high-spin
spectroscopy is the discovery of one-phonon, and possibly double-phonon,
excitation of the nuclear wobbling rotational bands.  In this talk,
we first review the properties of observed wobbling motions,
and discuss the failure and success of the possible interpretation in terms
of the simple rotor model.  Then, we further
present results of our microscopic study in Hf and Lu nuclei
by means of the theoretical framework, the cranked mean-field
and the random phase approximation.

\end{abstract}


\section{Introduction}
\label{Intro}

  In this talk we would like to discuss the nuclear wobbling motion.
The wobbling motion is a spinning motion of asymmetric top,
namely triaxial rigid-body.  Quite recently, rotational bands
associated with this motion have been identified in Lu nuclei\scite{Odeg},
and this discovery of the wobbling rotational bands is
one of the most exciting topics in the nuclear spectroscopy.
We must confess that we have been working on the wobbling motion
for rather long period.  In fact the talker(YRS)'s doctor thesis
is somewhat related to it.
So we are very regrettable that we have not been able to predict
the possible existence of them in the Lu isotopes
before the experiments.

  The reason why the wobbling motion is so exciting is that
it is related to a fundamental question: How does an atomic nucleus rotate
as a {\it three-dimensional} object?  Namely, the rotational motion
is neither uniform, nor the conventional one where
the axis of rotation coincides with one of principal axes.
Here we would like to stress
that most of the rotational bands, including the striking
high-spin 2:1 superdeformed bands, are supposed to be based
on the uniform rotation around the axis perpendicular to
the symmetry-axis of deformation,
so they are not genuine three-dimensional rotation.
Since the existence of the wobbling motion requires
the triaxial deformation, it also gives a rare chance to study the nuclear
mean-field with triaxial deformation,
which is very scarce near the ground state region.

  Recently, another type of exotic rotations, other than the usual
rotations around the perpendicular axis of axially symmetric nuclei,
have been also reported; that is the ``tilted axis rotation''
or ``magnetic rotation\scite{Fra}'', which is conceptually different from
the wobbling motion.  The wobbling motion is non-uniform rotation
and, just like the classical rigid body rotation,
the angular momentum vector is not parallel to
the rotational frequency vector,
while the tilted axis rotation is an uniform rotation
so that the two vectors are parallel with each other.
The typical electromagnetic transitions
associated with the wobbling excitations
are electric quadrupole (E2), while the magnetic dipole (M1)
transitions are very large in the tilted rotational bands.
Another important difference is that the tilted rotational band
appears as an isolated band (or a pair of bands in the case of recently
proposed ``chiral rotation/vibration''$^{\rcite{FM},\rcite{Staro}}$),
while the wobbling motion manifest itself
as a multi-rotational-band structure,
reflecting that the complex rotational motions
are composed of non-linear superposition of three rotations
around three principal axes of triaxially deformed body.

\begin{figure}[hbt]
\centerline{
\epsfxsize=110mm\epsffile{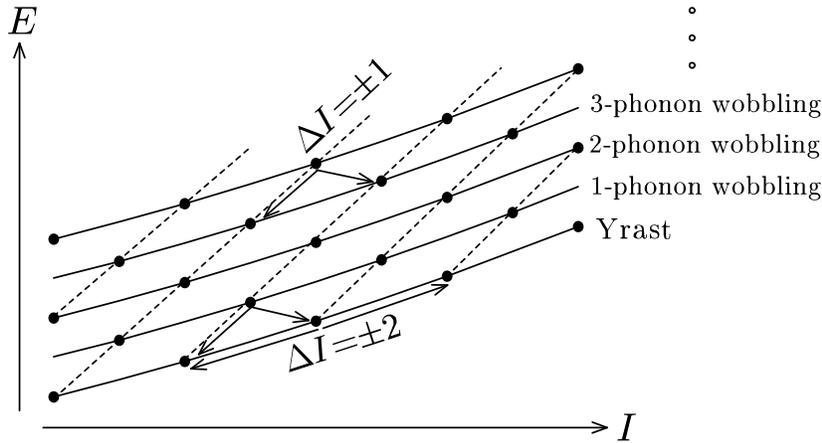}
}
\caption{\small \baselineskip 6pt
Schematic figure representing the multi-rotational-band
structure of the wobbling motion.
}
\label{fig:wobschem}
\end{figure}

  The band structure associated with the wobbling motion
is shown schematically in Fig.~\ref{fig:wobschem}.
The lowest band, namely the yrast band, corresponds to an uniform
rotation around the axis of largest moment of inertia.
One-phonon wobbling band is a rotational band with
a quantized wobbling phonon being excited
on top of the yrast band, which leads to a fluctuating motion
of the rotation axis with respect to the one in the yrast band.
Two-phonon wobbling band corresponds to
the band with two-wobbling quanta being excited on the yrast,
and the amplitude of fluctuation of the rotation axis is getting larger.
Three or more phonon bands are similar and based on multi-phonon excitations.
The characteristic of this band structure is that relatively strong
non-stretched E2 transitions connect the $n$-phonon and $(n-1)$-phonon bands:
The horizontal bands are usual rotational sequences with strong stretched
${\mit\Delta}I=\pm 2$ E2 transitions, while the vertical out-of-band
transitions between, e.g., one-phonon to vacuum yrast band are
${\mit\Delta}I=\pm 1$ E2 transitions, which are weaker than
the horizontal ones but much stronger than the usual
vibrational transitions.
The energy of vertical excitation $\hbar\omegaw$ is common
in all the $(n-1)$-phonon to $n$-phonon excitations and E2 transitions
between the $n$-phonon and $(n-2)$-phonon bands are prohibited
in a harmonic approximation.  This wobbling energy $\hbar\omegaw$
is given by the well known formula\scite{BM2}
in terms of three moments of inertia
around the principal axes of a rotating body,
which is discussed more closely in the following sections.

\section{Wobbling Motion: Observation and Simple Model Analysis}

\begin{figure}[hbt]
\centerline{
\epsfxsize=150mm\epsffile{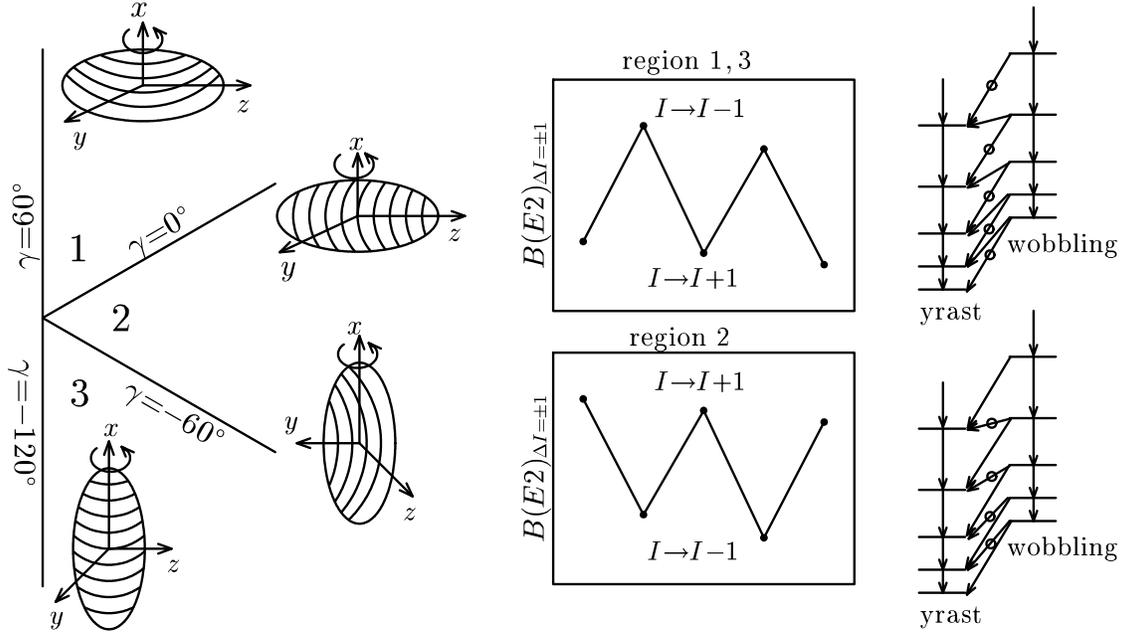}
}
\caption{\small \baselineskip 6pt
Schematic figure depicting the relation between the triaxial deformation
and the properties of the out-of-band E2 transition.
The shape corresponding to the triaxiality parameter
$\gamma$ (Lund convention) is shown relative to the main rotation axis,
which is chosen to the $x$-axis in the left panel.  In the middle,
the out-of-band $B$(E2) values with ${\mit\Delta}I=\pm 1$ are plotted
as functions of the angular momenta or the rotational frequency,
for which stronger transitions are marked in the right panel.
}
\label{fig:wobE2}
\end{figure}

  The first multi-rotational-band structure associated with
the wobbling motion have been observed
in $^{163}$Lu$^{\rcite{Odeg},\rcite{Jen}}$.
The low-spin structure in this nucleus is that of typical
well-deformed nucleus; there are many strongly-coupled and
aligned rotational bands.  At high-spin states, $I \gtsim 20\hbar$,
very regular rotational sequences invade into the yrast region,
whose moments of inertia are larger than the usual low-spin bands.
These sequences, totally four bands identified in  $^{163}$Lu,
are believed to be rotational bands based on
triaxial and largely deformed configurations,
which had been originally speculated
by calculations of the potential energy surface more than twenty years ago.
Nowadays, similar type of rotational bands are systematically
identified in this mass region, Lu and Hf nuclei, and called
the triaxial superdeformed (TSD) band.  Their triaxiality and deformation
are typically $\gamma \approx +20^\circ$
and $\epsilon_2 \approx 0.4$ (in the Lund convention,
see Fig.~\ref{fig:wobE2}), while those of the low-spin normal deformed states
are  $\gamma \approx 0^\circ$ and $\epsilon_2 \approx 0.2$.
The recent experimental progress in $^{163}$Lu is that the interband
transitions between TSD1 and TSD2 bands have been observed,
where the TSD1 is the yrast TSD band and
TSD2 is the first excited (one-phonon) band, and so they indicate clearly
the wobbling band structure mentioned in \S\ref{Intro}.
The measured out-of-band transitions is of $I$ to $I-1$,
and it has been confirmed to be mainly of E2 character.
The $B(E2)$ values\scite{Gor} are large
and can be nicely reproduced by the simple rotor model.
Moreover, the transitions
between TSD3 (two-phonon band) and TSD2, and between TSD3 and TSD1
have been also measured afterward\scite{Jen}.
Quite recently, the same band
structure have been observed in neighbouring Lu isotopes,
$^{165}$Lu\scite{Lu165} and $^{167}$Lu\scite{Lu167}.
In even-even Hf isotopes,
$^{168}$Hf\,\scite{Hf168} and $^{174}$Hf\,\scite{Hf174},
in the same mass region, excited TSD bands have been also measured,
although the linking transitions are not measured yet unfortunately.

  The interpretation of the observed band structure nicely fits into
the prediction first given by Bohr-Mottelson\scite{BM2} based on
the simple rotor model.  Let us recall the argument.
The rotor hamiltonian is composed of the three body-fixed
angular momenta ($J$'s) and three moments of inertia ($\cJ$'s)
around the principal axes in the body-fixed frame:
\beq
 H_\rot=\frac{J_x^2}{2\cJ_x}+ \frac{J_y^2}{2\cJ_y}+ \frac{J_z^2}{2\cJ_z}
 \equiv A_x J_x^2 +  A_y J_y^2 + A_z J_z^2.
\label{eq:Hrot}
\eeq
Assuming that $A_x < A_y < A_z$ $(\cJ_x > \cJ_y > \cJ_z)$, the yrast
state at given angular momentum $I$ is the uniform rotation
around the $x$-axis,
\beq
 E_I=\frac{\hbar^2I^2}{2\cJ_x}; \quad  \braketa{J_x} \approx \hbar I, \quad
  \braketa{J_y},\,\braketa{J_y} \approx \hbar \sqrt{I}.
\label{eq:Eyrast}
\eeq
Then the rotational frequency $\omega_\rot$ is defined, as usual,
by $\hbar\omega_\rot = dE_I/dI$.
In the excited state at given $I$ a nucleus rotate non-uniformly around
the axis which fluctuates around the main rotation axis ($x$-axis),
so that the fluctuating motion of the angular momentum vector
is described, in the harmonic approximation, by the following
normal mode creation operator:
\beq
 X^\dagger_{\rm w} = a \frac{iJ_y}{\sqrt{2I}}-b\frac{J_z}{\sqrt{2I}},
 \quad {\rm with}\quad [iJ_y, J_x] =J_x \approx I,
\label{eq:wobmode}
\eeq
where the amplitude $a$ and $b$ satisfy $ab=1$ due to the normalization
condition $[X_{\rm w},X^\dagger_{\rm w}]=1$, and they are
determined by diagonalizing the rotor hamiltonian (\ref{eq:Hrot});
$[H_\rot,X^\dagger_{\rm w}] =\hbar \omegaw X^\dagger_{\rm w}$.
The explicit calculation leads
$\sqrt{2}\,(A_y-A_x)b=\hbar\omegaw a/\sqrt{2I}$
and $\sqrt{2}\,(A_z-A_x)a=\hbar\omegaw b/\sqrt{2I}$,
thus $(\hbar\omegaw)^2=4I(A_y-A_x)(A_z-A_x)$, namely,
\beq
 \hbar\omegaw=I\sqrt{(1/\cJ_y-1/\cJ_x)((1/\cJ_z-1/\cJ_x)}
 =\hbar\omega_\rot
 \sqrt{\frac{(\cJ_x-\cJ_y)(\cJ_x-\cJ_z)}{\cJ_y\,\cJ_z}},
\label{eq:wobenergy}
\eeq
where $\omega_\rot=I/\cJ_x$ is the rotational frequency of
the yrast rotational band.
By using this eigen-mode (wobbling mode), the electric E2 transition
probabilities of both in-band transitions
in the yrast or the one-phonon wobbling bands
and out-of-band transitions between them can be also calculated.
The basic features are summarized in Fig.~\ref{fig:wobE2}.
As is usual, $B(E2)$ values are sensitive to the deformation,
especially in this case to the triaxiality.
Here it is to be noted that $0 \le \gamma \le 60^\circ$ is enough
for the triaxiality parameter $\gamma$ to specify the shape at zero spin
(no rotation).  At high-spin states, however, there is an axis of rotation
and there are two more regions of triaxial deformation
relative to the direction of the rotation axis ($x$-axis).
In the figure four possible rotation schemes with
axially symmetric deformation are shown, and there are three regions
of rotations with triaxial deformation in-between them.
As for the E2 transitions from the one-phonon wobbling band
to the yrast band, there are two possible transitions,
namely from $I$ to $I+1$ or from $I$ to $I-1$.  Which transition
is stronger is different in each region of triaxiality;
the $I$ to $I-1$ transition is stronger in the region 1 and 3,
i.e. $0^\circ < \gamma < 60^\circ$ or $-120^\circ < \gamma < -60^\circ$,
while the $I$ to $I+1$ transition is stronger in the region 2, i.e.
i.e. $-60^\circ < \gamma < 0^\circ$.

  Taking into account the basic properties of the wobbling motion
in the rotor model, the key quantities are 1) three moments of inertia,
$\cJ_x$, $\cJ_y$, $\cJ_z$, and 2) the triaxial deformation, especially
the sign of the triaxiality parameter, $\gamma$;
both are related in some way.
In order for the existence of the wobbling motion, $\cJ_x$ should be
the largest (see Eq.~(\ref{eq:wobenergy})), where $x$-axis is
the axis of main rotation of the yrast TSD band.
On the other, the $B(E2)$ data suggests positive $\gamma$ shape,
$0^\circ < \gamma < 60^\circ$,
since only the $I$ to $I-1$ transitions are observed\footnote[2]{
 The possibility of $-120^\circ < \gamma < -60^\circ$, the region 3
 in Fig.~\ref{fig:wobE2}, cannot be excluded, but it is
 very unlikely from the spectra of the yrast TSD band.},
which is also consistent to the calculated position of minimum
corresponding to the TSD band in the potential energy surface,
i.e. $\gamma({\rm PES}) \approx +20^\circ$.
It is, however, stressed that
the irrotational moments of inertia, which are commonly used
in the rotor model, require $\cJ_y > \cJ_x > \cJ_z$ at
positive $\gamma$ ($0^\circ < \gamma < 60^\circ$), and clearly contradict
the existence of the wobbling motion
($\hbar\omegaw$ in Eq.~(\ref{eq:wobenergy})  becomes imaginary).
The classical rigid moments of inertia satisfies the condition,
in fact $\cJ_x > \cJ_y > \cJ_z$ at positive $\gamma$, but they do not
meet the basic quantum mechanics criteria that the rotation should
not occur around the symmetry axis,
in contrast to the irrotational inertia.

  One of the other observed features of the wobbling motion
in Lu isotopes is that the wobbling excitation energy $\hbar\omegaw$
decreases as a function of the spin $I$ or the rotational frequency
$\omega_\rot$ (see the left panel of Fig.~\ref{fig:RPAcal4} shown later).
This trend seems common to all the observed cases in Lu isotopes,
but completely opposite to that predicted by the simple rotor model;
$\hbar\omegaw$ in Eq.~(\ref{eq:wobenergy}) is proportional
to the rotational frequency if three moments of inertia are assumed
to be constant.  Therefore, the $\omega_\rot$-dependence
of the wobbling energy requires that three moments of inertia
should depend on $\omega_\rot$ in such a way to decrease $\hbar\omegaw$.
Another interesting feature to be pointed out is the magnitude
of out-of-band $B(E2)^{I\rightarrow I-1}_\out$, which amounts to
100 or more Weisskopf units.  Note that the in-band
$B(E2)^{I\rightarrow I-2}_\in$ value is larger and about 500 or more
Weisskopf units in TSD bands, which are consistent with
the calculated deformation parameters
($\epsilon_2 \approx 0.4, \gamma \approx +20^\circ$) by the potential
energy surface, and so the out-of-band transitions is about 20\%
of the in-band transitions.
These transitions are extremely strong
and suggests that both are of rotational origin,
which are nicely reproduced by the simple rotor model.
As examples of strong out-of-band E2 transitions,
those between the ground state band and the collective
$\beta$- or $\gamma$-vibrational band are known in well-deformed nuclei.
Their $B(E2)$ values are typically about 5 -- 8 Weisskopf units,
while the typical in-band E2 transition probabilities
of normal deformed nuclei are 100 -- 200 Weisskopf units.

  In the Lu isotopes the odd proton particle exists in addition
to the simple rotor. Therefore one has to consider the more elaborated
particle-rotor coupling model\scite{Ham}.
The essential features discussed above are not changed as long as
three moments of inertia satisfying $\cJ_x > \cJ_y > \cJ_z$ are used,
although the presence of the odd particle makes
the rotational spectra more complex.

\section{Microscopic RPA Model}
\label{RPA}

  Considering the observed features of the wobbling motion discussed
in the previous section, the simple rotor model fails; namely
one cannot use the irrotational moments of inertia and
the dependence of three moments of inertia on the rotational frequency
should be taken into account.  Since we do not know what kind of
moments of inertia should be used {\it a priori}, we have to
calculate three moments of inertia, which requires a microscopic
framework to study the wobbling motions.  Such a framework
were proposed by Marshalek\scite{Mar},
and examined in details in some realistic cases in Ref.~\rcite{SM}.

\begin{figure}[hbt]
\centerline{
\epsfxsize=130mm\epsffile{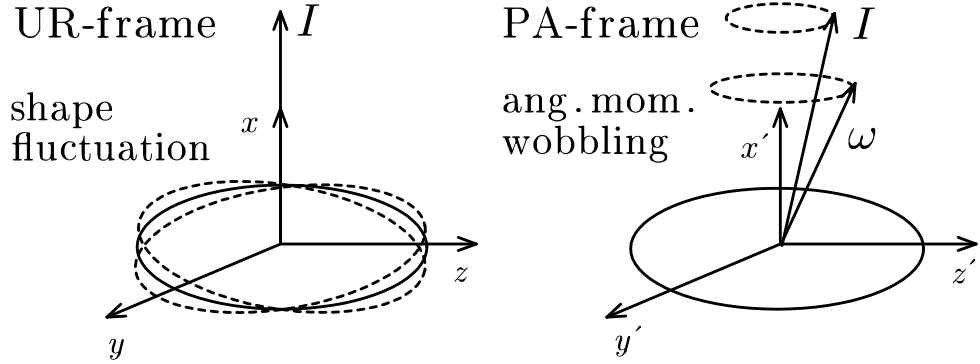}
}
\caption{\small \baselineskip 6pt
Schematic figure depicting the two dynamical pictures
in the uniformly rotating frame (UR frame) and
the principal axis frame (PA frame).
}
\label{fig:wobRPA}
\end{figure}

  The theory is based on the random phase approximation (RPA)
on top of the cranked mean-field, which is used to describe
the uniform rotation of the yrast (vacuum) rotational band
by the stationary mean-field hamiltonian, $h'=h_\defo -\omega_\rot J_x$.
In the one-phonon wobbling band the fluctuating motion is not so large,
and the small amplitude approximation of
time-dependent mean-field around $h'$ can be used,
which results in the RPA eigen-mode equation.
Thus, with using the $QQ$ type force as a residual interaction,
the $n$-th eigen-energy $\hbar\omega_n$ and
eigen-mode creation operator $X_n^\dagger=\sum_{\alpha\beta}\left[
 \psi(\alpha\beta)a^\dagger_\alpha a^\dagger_\beta -
 \phi(\alpha\beta)a_\beta a_\alpha \right]$
as a superposition of two-quasiparticle excitations
can be calculated.  If the $n$-th mode is identified as a wobbling
motion, $X_n^\dagger$ and $\hbar\omega_n$ correspond to
$X^\dagger_{\rm w}$ and $\hbar\omegaw$ in Eqs.~(\ref{eq:wobmode}),
(\ref{eq:wobenergy}) in the simple rotor model.
The $QQ$ type force contains the quadrupole tensor,
$Q_{ij}=\sqrt{\frac{15}{4\pi}}\sum_{a=1}^{A}\left[
 x_i x_j - \frac{1}{3} r^2 \delta_{ij} \right]_a$
($i,j=x,y,z$, in the Cartesian representation),
but, because of the symmetry such that the wobbling excitation
changes the angular momentum by $\pm 1$ unit, only the two components,
$Q_y \equiv -Q_{zx}$ and $Q_z \equiv i\,Q_{xy}$ are responsible
for dynamical time-dependence\footnote[2]{
 If is used the usual spherical tensor representation with
 classification by the signature quantum number,
 $Q_{2K}^{(\pm)}$, then $Q_y=Q_{21}^{(-)}$ and $Q_z=Q_{22}^{(-)}$, i.e.
 they transfer signature by one unit.}.
Thus the time-dependent mean-field is
\beq
 h_{\rm UR}(t)=h_\defo - \omega_\rot J_x
 - \kappa_y {\cal Q}_y(t) Q_y- \kappa_z {\cal Q}_z(t) Q_z,
\label{eq:hUR}
\eeq
where $\kappa_{y,z}$ are the $QQ$ type force strengths, and
${\cal Q}_{y,z}(t)=\braketc{t}{Q_{y,z}}{t}$ describe
the time-dependence of the relevant quadrupole components.
The subscript UR is attached because this time-dependent hamiltonian
is defined in the uniformly rotating (UR) frame, where the rotation axis
is pointing to the main rotation axis ($x$-axis) of the vacuum band.
The wobbling excitation on it induces the shape fluctuation of
the non-diagonal quadrupole tensor,
${\cal Q}_y(t)$ and ${\cal Q}_z(t)$, in the mean-field,
and the out-of-band E2 transition probabilities are
calculated by
\beq
 B(E2)_\out^{I\rightarrow I\mp 1}
 = {\textstyle\frac{1}{2}} |{\cal Q}^{(E)}_y(n)\pm{\cal Q}^{(E)}_z(n)|^2,
\label{eq:BE2}
\eeq
where ${\cal Q}^{(E)}_{y,z}(n)$ are the electric (proton) part of
tensor calculated with only the $n$-th eigen-mode being excited.
In order to recover the wobbling picture of the angular momentum
fluctuation, the time-dependent non-unitary transformation
to the principal axis (PA) frame should be performed by requiring
that the non-diagonal part of quadrupole tensor should vanish.
Then the time-dependent mean-field in the PA frame is now
\beq
 h_{\rm PA}(t)=h_\defo - \omega_x(t) J_x
 - \omega_y(t) J_y - \omega_z(t) J_z,
\label{eq:hPA}
\eeq
where $\omega_x(t) \approx \omega_\rot$ in the small amplitude limit,
and $\omega_{y,z}(t)$ being related to ${\cal Q}_{y,z}(t)$ describe
the fluctuation of the angular frequency vector.  In this frame,
the fluctuation of the angular momentum vector naturally arises,
$\braketc{t}{J_{y,z}}{t}$: Then the three RPA moments of inertia are
introduced through
\beq
 {\cal J}_x\equiv\braketa{J_x}/\omega_\rot, \quad
 {\cal J}_{y,z}(n)\equiv J_{y,z}(n)/\omega_{y,z}(n),
\label{eq:Tmom}
\eeq
where the frequencies $\omega_{y,z}(n)$ and the expectation values
$J_{y,z}(n)$ are calculated with only the $n$-th eigen-mode being excited.
What Marshalek found is
that using these three moments of inertia the RPA phonon energy
can be expressed in the same way as in Eq.~(\ref{eq:wobenergy})
in the simple rotor model, namely the wobbling energy equation
is equivalent to the RPA eigen-mode equation.
The dynamical pictures of the wobbling motion in the UR and PA frames
are shown schematically in Fig.~\ref{fig:wobRPA}.

  In this way, the wobbling phonon energy, the $B(E2)$ values,
and the three moments of inertia can be calculated
in a microscopic framework without ambiguity.
We would like to stress that the $QQ$ type force strengths $\kappa_{y,z}$
in Eq.~(\ref{eq:hUR}) are not free parameters but fixed by
the requirement of the decoupling of the Nambu-Goldstone modes in the RPA.
Therefore, there is no adjustable parameters once the mean-field
parameters are fixed selfconsistently.
It should also be noticed that the number of RPA eigen-modes
are that of independent two-quasiparticle states, but most of the
solutions do not have a proper wobbling property.  For example,
the defined ${\cal J}_{y,z}(n)$ can take negative values,
or the fluctuation amplitude of the angular momentum vector
is too small if the obtained E2 amplitudes
${\cal Q}_y(n)$ and ${\cal Q}_z(n)$ are not collective;
such solutions are not wobbling mode at all.
In fact, the RPA solution which can be interpreted as a wobbling motion
do not always appear: Some condition on the mean-field is necessary.

\begin{figure}[hbt]
\centerline{
\epsfxsize=65mm\epsffile{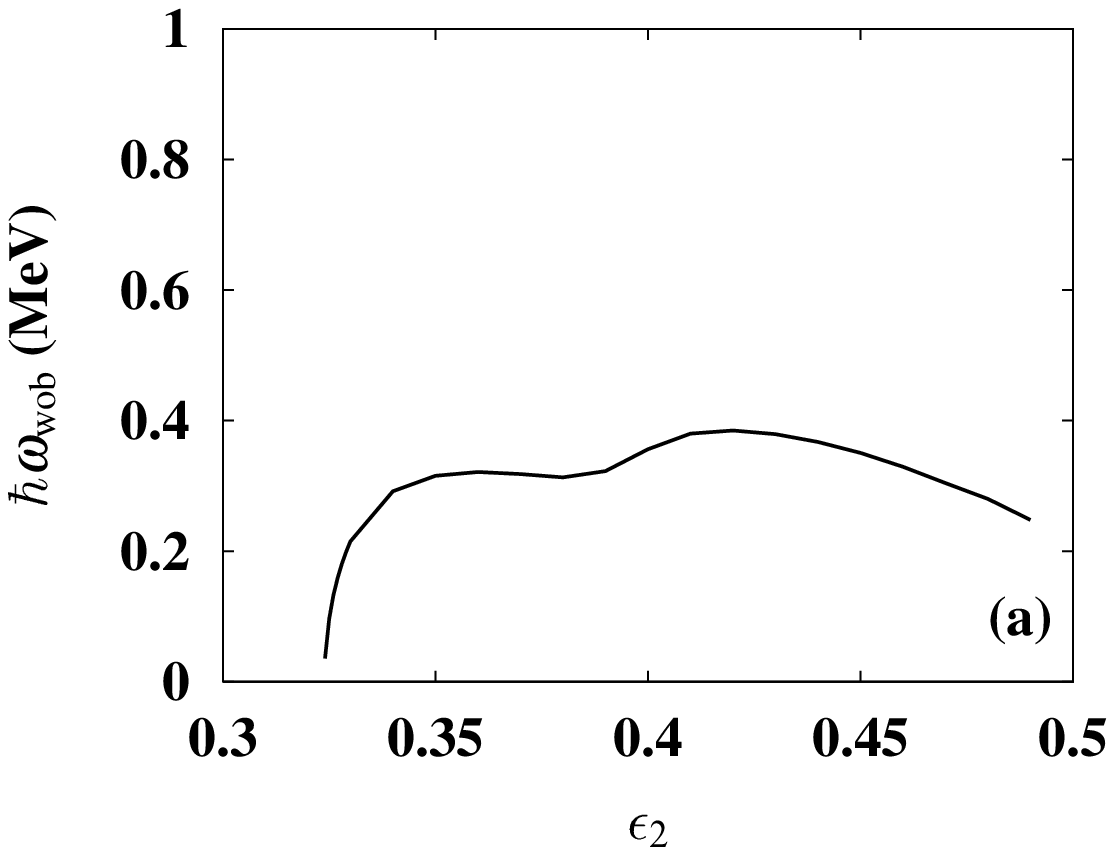}
\epsfxsize=65mm\epsffile{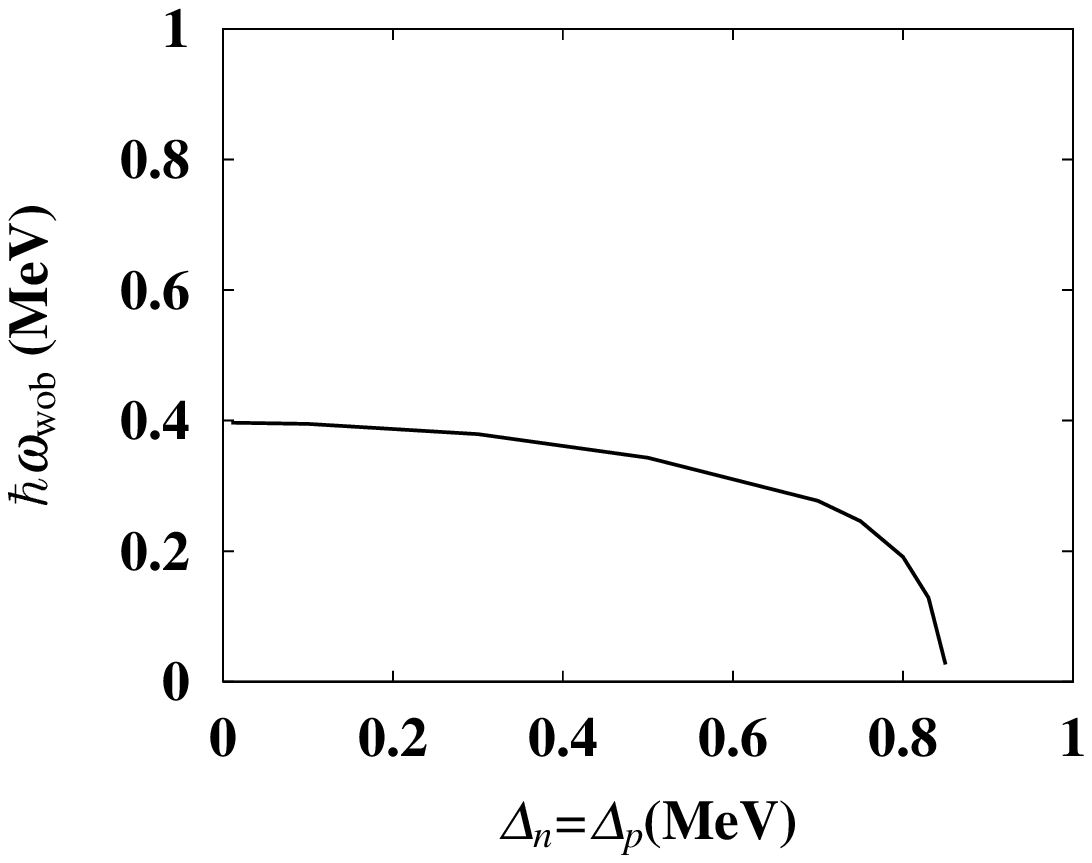}
}
\caption{\small \baselineskip 6pt
Wobbling excitation energy calculated as a function
of the deformation parameter $\epsilon_2$ (left), and
of the static pairing gap parameter ${\mit\Delta}_{\rm n,p}$ (right),
in an even-even nucleus $^{168}$Hf. Taken from Ref.~\protect\rcite{MSMb}.
}
\label{fig:RPAcal1}
\end{figure}

\begin{figure}[hbt]
\centerline{
\epsfxsize=65mm\epsffile{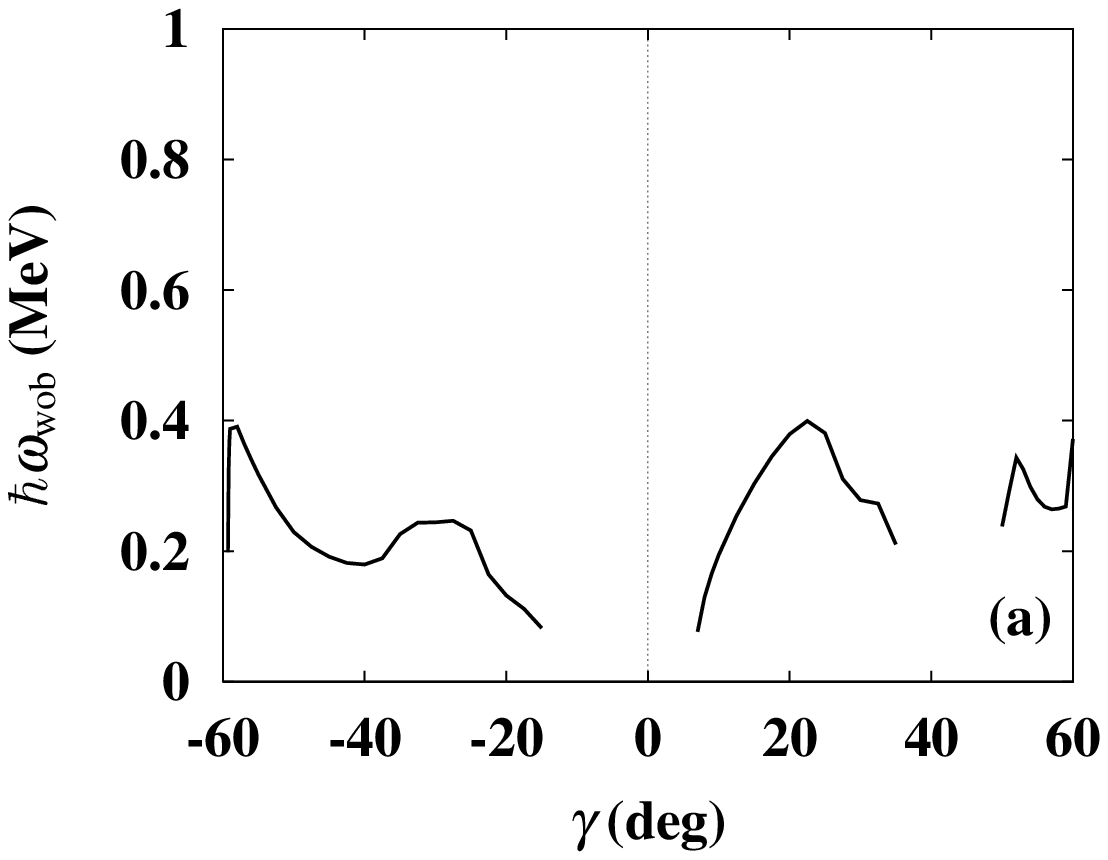}
\epsfxsize=65mm\epsffile{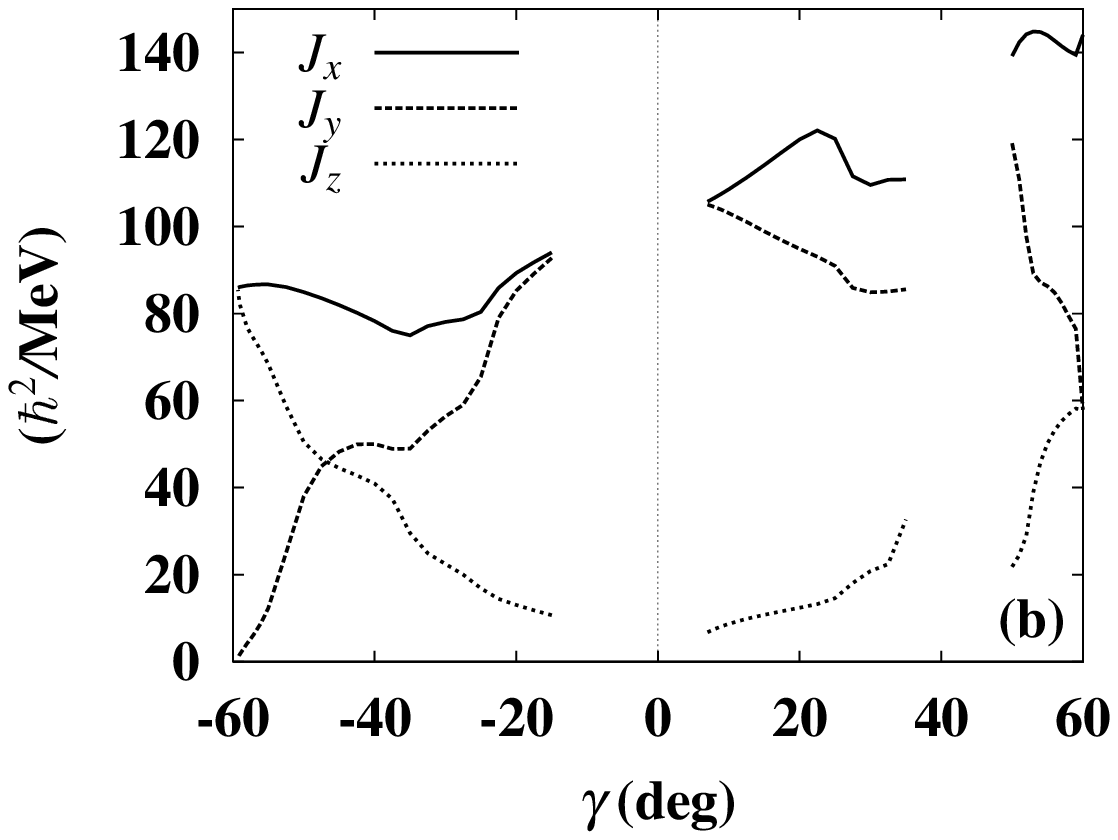}
}
\caption{\small \baselineskip 6pt
Wobbling excitation energy (left) and three moments of
inertia (right) as functions
of the triaxiality parameter $\gamma$,
in an even-even nucleus $^{168}$Hf. Taken from Ref.~\protect\rcite{MSMb}.
}
\label{fig:RPAcal2}
\end{figure}

\begin{figure}[hbt]
\centerline{
\epsfxsize=65mm\epsffile{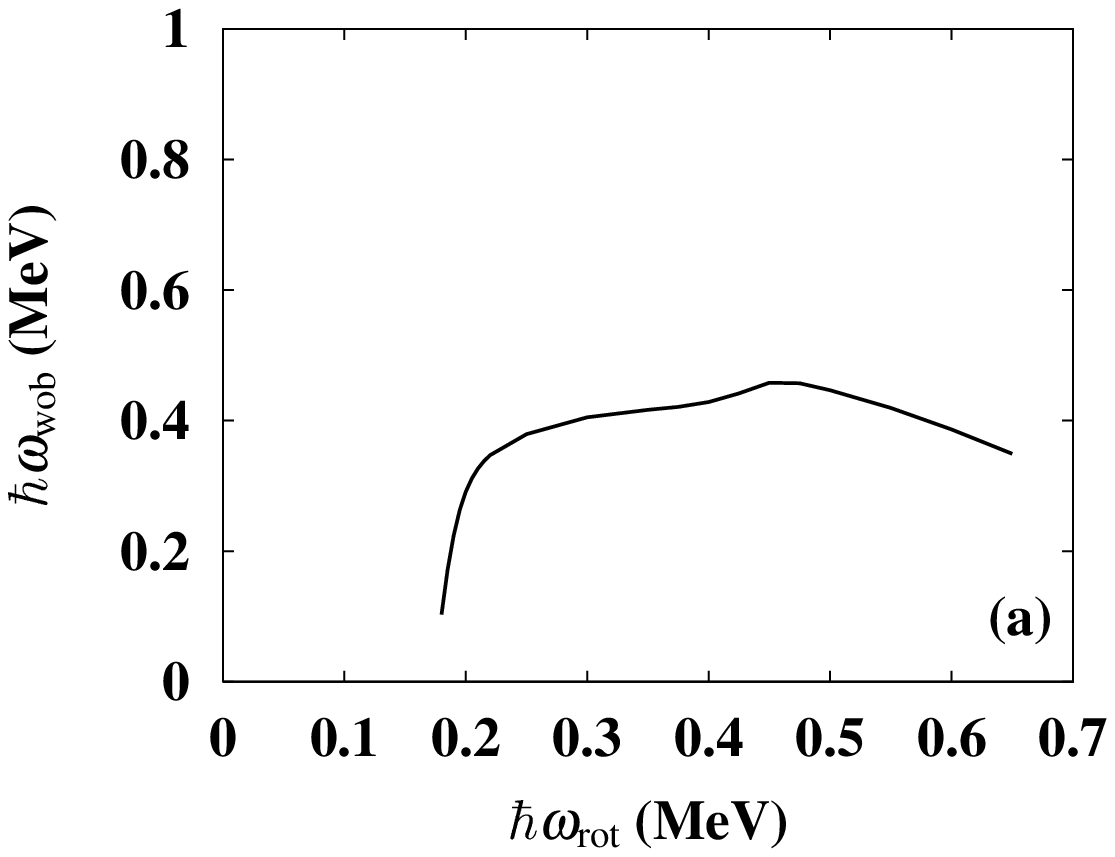}
\epsfxsize=65mm\epsffile{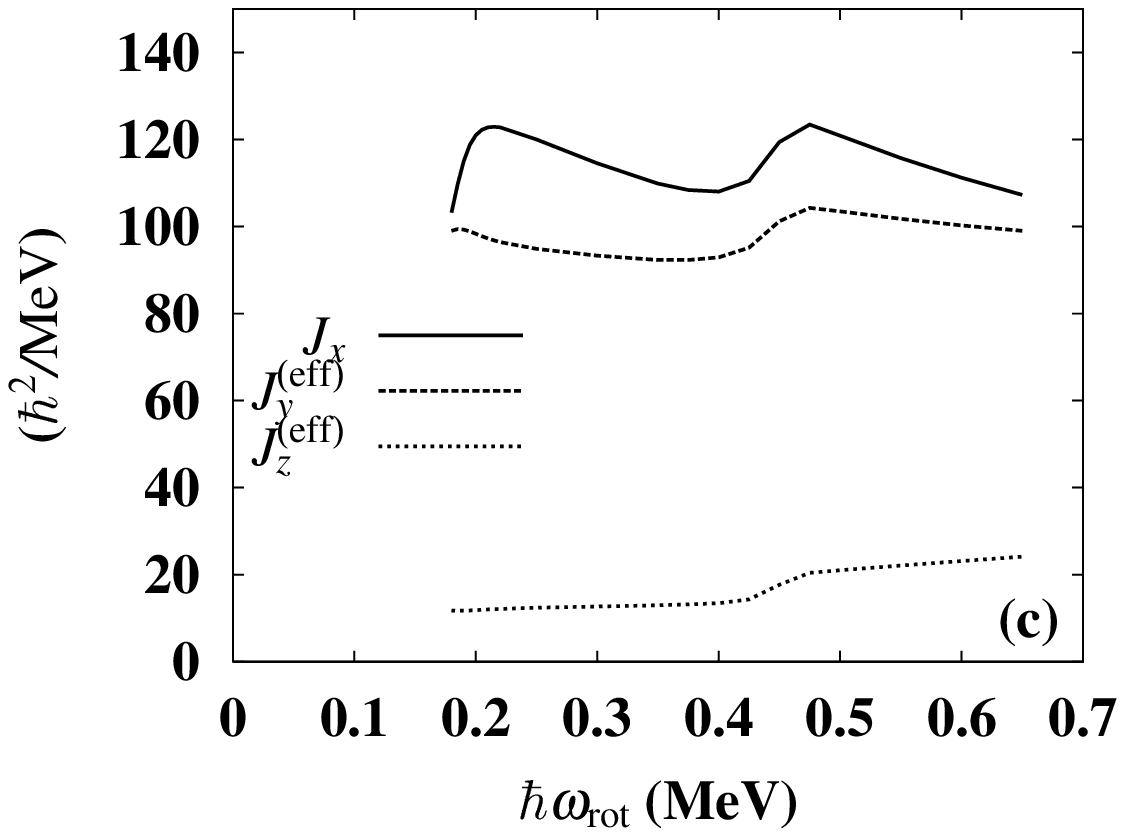}
}
\caption{\small \baselineskip 6pt
Wobbling excitation energy (left) and three RPA moments of
inertia (right) as functions of the rotational frequency $\omega_\rot$,
in an even-even nucleus $^{168}$Hf. Taken from Ref.~\protect\rcite{MSMb}.
}
\label{fig:RPAcal3}
\end{figure}

  In Figs.~\ref{fig:RPAcal1} and \ref{fig:RPAcal2} we show an example
depicting the dependences
of the wobbling energy $\hbar\omegaw$ on various mean-field parameters,
$\epsilon_2$, $\gamma$ and pairing gap ${\mit\Delta}_{\rm n,p}$ calculated
at $\hbar\omega_\rot=0.3$ MeV in a even-even nucleus $^{168}$Hf.
The collective wobbling solution indeed exists around the expected
values of parameters $\epsilon_2 \approx 0.4$ and $\gamma \approx +20^\circ$,
and it is stable against the change of the pairing gap parameters,
which are supposed to be small (${\mit\Delta} \ltsim 0.5$ MeV)
in the TSD bands. It should be noted that the wobbling mode
becomes softer ($\hbar\omegaw$ decreases) as ${\mit\Delta}$ increases,
which is opposite behaviour to the case of the conventional
collective vibrational modes,
and may indicate that it is of rotational character.
The three RPA moments of inertia are also shown in Fig.~\ref{fig:RPAcal2},
from which it is clear that they are neither irrotational nor
rigid-body like.  We further show the wobbling energy
and RPA moments of inertia as functions 
of the rotational frequency $\omega_\rot$ in Fig.~\ref{fig:RPAcal3};
unfortunately the wobbling motion is not established yet in this nucleus.
Here the mean-field parameters are fixed for simplicity;
$\epsilon_2=0.43$, $\gamma=+20^\circ$,
and ${\mit\Delta}_{\rm n}={\mit\Delta}_{\rm p}=0.3$ MeV.
In the microscopic RPA calculation the $\omega_\rot$-dependence
naturally arises as a result of cranking prescription
of the quasiparticle orbits, and the wobbling energy is not simply
proportional to the rotational frequency.

  It is known that the microscopically calculated $\gamma$-dependence
of three moments of inertia at zero rotational frequency,
e.g. by the Inglis cranking formula,
look very similar to the irrotational inertia.
If that is the case, why does the wobbling solution appear
in our microscopic RPA calculations?
The reason is the following: The $\cJ_x$ inertia in the RPA formalism
in Eq.~(\ref{eq:Tmom}) is that of kinematic moment of inertia,
so that the alignments of quasiparticle orbits contribute to it.
Actually the occupation of the high-$j$ proton $i_{13/2}$ quasiparticle
is essential to generate a minimum at the positive $\gamma$ shape
for the TSD bands.  As is seen in Fig.~\ref{fig:RPAcal3},
the alignment of two $\pi i_{13/2}$ quasiparticles occurs around
$\hbar\omega_\rot \approx 0.2$ MeV, which suddenly increases $\cJ_x$.
Because of this effect the condition $\cJ_x > \cJ_y(n) > \cJ_z(n)$
is satisfied and the wobbling solution appears.
Thus the increase of $\cJ_x$ due to the quasiparticle alignments
is crucial for the appearance of the wobbling motion
in our RPA calculations; see Ref.~\rcite{MSMa} and \rcite{MSMb} for details.

\begin{figure}[hbt]
\centerline{
\epsfxsize=65mm\epsffile{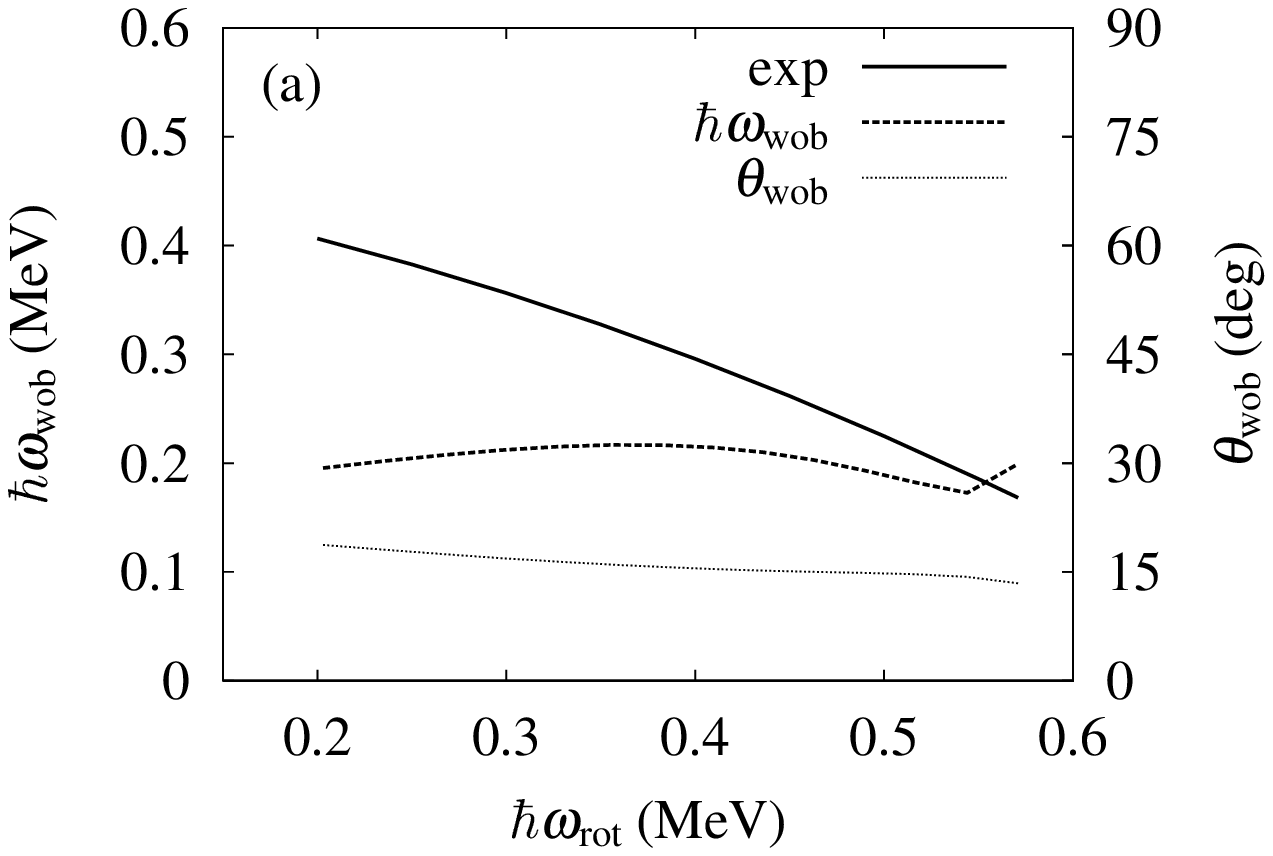} \hskip 10mm
\epsfxsize=65mm\epsffile{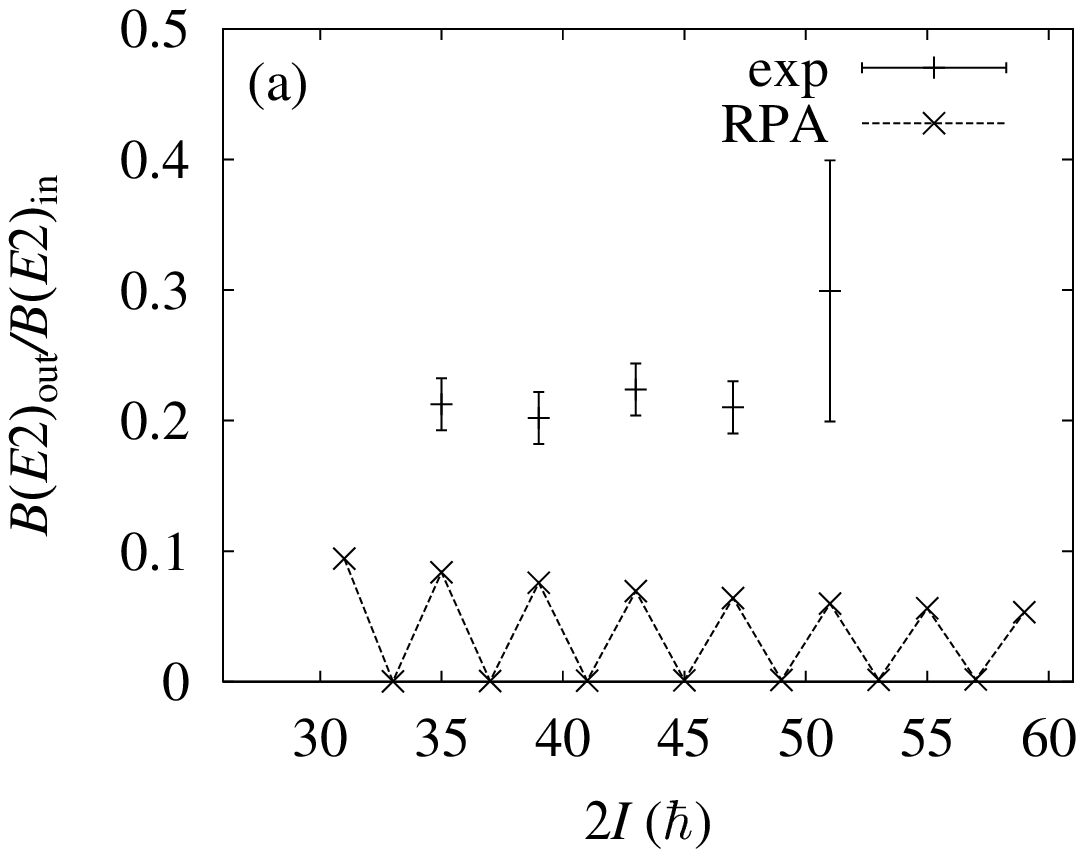}
}
\caption{\small \baselineskip 6pt
Comparison of the calculated wobbling energy (left) and
the $B(E2)$ value (right) with experimental data in $^{163}$Lu nucleus.
In the left panel, the wobbling amplitude
$\theta_{\rm w} \equiv \tan^{-1}{(\sqrt{J_y(n)^2+J_z(n)^2}/\braketa{J_x})}$
is also shown, see Ref.~\protect\rcite{MSMa} for details.
}
\label{fig:RPAcal4}
\end{figure}

  Now we compare the calculated results with experimental data
for $^{163}$Lu in Fig.~\ref{fig:RPAcal4}, where, again, all the
mean-field parameters are fixed.  The calculated
wobbling energy $\hbar\omegaw$ is smaller than the experimental
data and stays almost constant against the rotational frequency $\omega_\rot$,
while, as already mentioned, the experimental wobbling energy
decreases with $\omega_\rot$. Thus, the result of calculation is not
very successful to reproduce the detailed $\omega_\rot$-dependence.
This requires that the change of the mean-field parameters as functions
of $\omega_\rot$ should be considered.
Even more problematic is the out-of-band $B(E2)$ values,
which are about 2--3 times smaller than the experimentally measured values.
Here we would like to recall that the measured $B(E2)$ is very large,
more than 100 Weisskopf units: Although the calculated $B(E2)$ values
are extremely large compare to those in the case of
usual collective vibrations, it is not enough to reproduce those
of the wobbling mode.  Considering the sum-rule like argument
in Ref.~\rcite{MSMb}, we feel it is very difficult for the microscopic
RPA theory to understand that the macroscopic rotor limit of
the out-of-band $B(E2)$ value is almost reached in the actual nucleus.

\section{Summary and Discussions}

  In this talk, recently identified nuclear wobbling motions
in the Lu and Hf region are reviewed and discussed
from the microscopic view point.  The original picture of the wobbling motion
is based on the simple rotor model.  It is, however, apparent that
the atomic nucleus is not macroscopic object like a rotor.
Therefore, the observed properties of the wobbling motion
cannot be always understood by the rotor model, and one has to
invoke more fundamental microscopic theories.
We summarize the points of our investigations up to now,
which have been done by means of
the microscopic RPA framework in the previous section, as follows:
\begin{enumerate}
\renewcommand{\labelenumi}{(\arabic{enumi})}

\item By suitable choice of the mean-field parameters,
i.e. large enough $\epsilon_2$ and positive $\gamma\approx +20^\circ$,
we have found that low-energy wobbling solutions appear naturally
among the RPA eigen-modes.  The wobbling solution is insensitive
to the pairing gap parameters; the eigen-energy decreases as
the gap increases, which is a completely opposite behaviour
compared with the case of low-lying collective vibrations.

\item The proton $i_{13/2}$ quasiparticle alignments are crucial
to obtain the condition $\cJ_x > \cJ_y(n) > \cJ_z(n)$ for
the RPA moments of inertia, which is required for the existence
of the associated wobbling mode.  It is consistent with the fact that
the occupation of the same proton $i_{13/2}$ quasiparticle is necessary
for the positive $\gamma$ TSD shape to be minimum
in the potential energy surface calculations.

\item The detailed rotational frequency dependence of the wobbling
excitation energy could not be reproduced in the present RPA calculation,
although it is much improved compared with the result of
the simple rotor model with fixed moments of inertia,
which gives completely opposite dependence to the experimental data.
All the mean-field parameters are assumed to be constant for simplicity
in our calculations.  This result suggests that change of the mean-field
against the rotational frequency should be properly taken into account.

\item  A severe problem of the RPA calculations is that the out-of-band
$B(E2)$ values are smaller by about factor two or three than
the experimentally measured values.  Although the obtained RPA solution
is extremely collective, the collectivity (enhancement of $B(E2)$)
is not enough.  This poses an important future challenge for
the microscopic theory like RPA.

\end{enumerate}

\begin{figure}[hbt]
\centerline{
\epsfxsize=75mm\epsffile{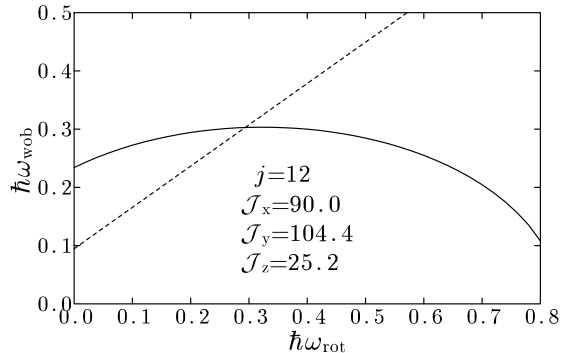}
}
\caption{\small \baselineskip 6pt
An example of the wobbling excitation energy
in the modified rotor hamiltonian (\protect\ref{eq:newHrot})
as a function of $\hbar\omega_\rot=(I-j)/\cJ_x$.
The (rather arbitrarily) chosen values of
the quasiparticle alignment $j$ and three inertia $\cJ_y > \cJ_x > \cJ_z$
are shown in the figure.  The dashed line is
the original wobbling energy (\protect\ref{eq:wobenergy})
calculated by replacing $\cJ_x$ and $\cJ_y$
(so-called $\gamma$-reversed inertia\protect\scite{Ham}).
}
\label{fig:modwob}
\end{figure}

  Finally we would like to discuss a possible explanation of
the calculated dependence of the wobbling energy $\hbar\omegaw$ on
the rotational frequency $\omega_\rot$ (or spin $I$), i.e.
increasing in the lower frequency and decreasing in the higher frequency,
which are shown in Figs.~\ref{fig:RPAcal3} and \ref{fig:RPAcal4}
(see Ref.~\rcite{MSMb} for more examples of calculation).
It is based on a rotor model with a modification to include
the effect of quasiparticle alignments,
which is investigated in Ref.~\rcite{Ham}, but the possibility
is not thoroughly explored\footnote[2]{The following argument
results from the discussion with Stefan Frauendorf.}.
The hamiltonian is the same as Eq.~(\ref{eq:Hrot}) except that
the $J_x$ is now replaced to $(J_x -j)$, where $j$ is a constant and
corresponds to the aligned angular momentum of quasiparticle(s).
Expanding the $(J_x-j)^2$ term, a linear term
$-jJ_x/\cJ_x$ in $J_x$ appears.
Treating this term by $J_x =\left[I^2-(J_y^2+J_z^2)\right]^{\frac{1}{2}}
\approx I - (J_y^2+J_z^2)/2I$ in the high-spin limit,
the modified rotor hamiltonian and the yrast energy are given by
\beq
 {\tilde H}_\rot  \approx
 \frac{J_x^2}{2\cJ_x}+ \frac{J_y^2}{2{\tilde \cJ}_y(I)}
   + \frac{J_z^2}{2{\tilde \cJ}_z(I)}
   + \left(\frac{j^2-2Ij}{2\cJ_x}\right), \quad
   E_I=\frac{(I-j)^2}{2\cJ_x},
\label{eq:newHrot}
\eeq
and the rotational frequency is now given by $\hbar\omega_\rot=(I-j)/\cJ_x$.
The last term in ${\tilde H}_\rot$ is constant, and
\beq
 {\tilde \cJ}_{y,z}(I)=\cJ_{y,z}\,
 \left[ 1+\frac{j\cJ_{y,z}}{I\cJ_x} \right]^{-1}
\label{eq:mTmom}
\eeq
are the modified $I$-dependent moments of inertia.
Namely, the effect of the quasiparticle alignments
makes ${\tilde \cJ}_{y,z}(I)$ inertia smaller.
Since the diagonalization of Eq.~(\ref{eq:newHrot}) in terms of
the wobbling mode (\ref{eq:wobmode}) is the same if $\cJ_{y,z}$ are
replaced by ${\tilde \cJ}_{y,z}(I)$, the wobbling mode appears
in the spin range, $j < I < I_{\rm crit}\equiv j[1-\cJ_x/\cJ_y]^{-1}$,
because the condition
$\cJ_x > {\tilde \cJ}_y(I) > {\tilde \cJ}_z(I)$ is satisfied
in $I < I_{\rm crit}$ ($j< I$ is required for $\hbar\omega_\rot=dE_I/dI > 0$),
even when the original three inertia are 
$\cJ_y > \cJ_x > \cJ_z$ like the irrotational inertia
at the positive $\gamma$ shape.  An example of the wobbling
excitation energy calculated with this modified ${\tilde H}_\rot$
is shown in Fig.~\ref{fig:modwob}.
Even though the original three inertia are constants, a similar
$\omega_\rot$-dependence to those in the realistic RPA calculations
emerges because of the presence of the quasiparticle alignments.
It is, however, mentioned that
the appearance mechanism of the wobbling motion is somewhat
different from the interpretation of the RPA calculation given in \S\ref{RPA}:
$\cJ_x$ is increased by the alignments in the RPA,
while $\cJ_{y,z}$ are decreased in this modified rotor model,
although the quasiparticle alignments play an essential role
in both explanations.  At this stage we are not sure that
this model serves as a possible model explaining the results
of the microscopic RPA calculations.  But we hope that this kind
of interpretation of the calculated results by an intuitive model
deepens insight into the appearance mechanism of the wobbling motion,
and gives some clues for improving the microscopic framework.

\vskip 5mm
\leftline{\Large\bf Acknowledgments}
\vskip 3mm

Discussion with Stefan Frauendorf at ECT* in Trento is greatly appreciated.
This work was supported in part by the Grant-in-Aid for scientific
research from the Japan Ministry of Education, Science and Culture
(Grant Nos. 13640281 and 14540269).

\vskip 5mm
\leftline{\Large\bf References}
\vskip 2mm
\begin{list}{}{
   \setlength{\itemindent}{0 pt}
   \setlength{\leftmargin}{20 pt}
   \setlength{\labelwidth}{16 pt}
   \setlength{\topsep}{0 pt}
   \setlength{\parsep}{0 pt}
   \setlength{\itemsep}{2 pt}
   \newcounter{myref}
   \setcounter{myref}{0}
   \refstepcounter{myref}
 \def \bibref#1{\item[\themyref )] \label{#1} \refstepcounter{myref}}
}


\bibref{Odeg}
 S.~W.~\O deg\aa rd et al., Phys. Rev. Lett. {\bf 85} (2001), 5866;
 D.~R.~Jensen et al., Nucl. Phys. {\bf A703} (2002), 2.

\bibref{Fra}
 S.~Frauendorf, Rev. Mod. Phys. {\bf 73} (2001), 463.

\bibref{FM}
 S.~Frauendorf and J.~Meng, Nucl. Phys. {\bf A617} (1997), 131.

\bibref{Staro}
 K.~Starosta, et al., Phys. Rev. Lett. {\bf 86} (2001), 971;
 T.~Koike, et al., Phys. Rev. {\bf C67} (2003), 044319. 

\bibref{BM2}
 A.~Bohr and B.~R.~Mottelson, {\it Nuclear Structure}, Vol.~II,
 Benjamin, New York, 1975.

\bibref{Jen}
 D.~R.~Jensen et al., Phys. Rev. Lett. {\bf 89} (2002), 140523;
 D.~R.~Jensen et al., Eur. Phys. J. {\bf A19} (2004), 173.

\bibref{Gor}
 A.~G{\"o}gen, et al., Phys. Rev. {\bf C69} (2004), 031301(R).

\bibref{Lu165}
 G.~Sch{\"o}nwa{\ss}er, et al., Phys. Lett. {\bf B552} (2003), 9.

\bibref{Lu167}
 H.~Amro, et al., Phys. Lett. {\bf B553} (2003), 197.

\bibref{Hf168}
 H.~Amro, et al., Phys. Lett. {\bf B506} (2001), 39.

\bibref{Hf174}
 M.~K.~Djongolov, et al., Phys. Lett. {\bf B560} (2003), 24.

\bibref{Ham}
 I.~Hamamoto, Phys. Rev. {\bf C65} (2002), 044305;
 I.~Hamamoto and G.~B.~Hageman, Phys. Rev. {\bf C67} (2003), 014319.

\bibref{Mar}
 E.~Marshalek, Nucl. Phys. {\bf A331} (1979), 429.

\bibref{SM}
 M.~Matsuzaki, Nucl. Phys. {\bf A509} (1990), 269;
 Y.~R.~Shimizu and M.~Matsuzaki, Nucl. Phys. {\bf A588} (1995), 559.

\bibref{MSMa}
 M.~Matsuzaki, Y.~R.~Shimizu, and K.~Matsuyanagi,
 Phys. Rev. {\bf C65} (2002), 041303(R).

\bibref{MSMb}
 M.~Matsuzaki, Y.~R.~Shimizu, and K.~Matsuyanagi,
 Phys. Rev. {\bf C69} (2004), 034325.

\end{list}
%

\vfill
\end{document}